\documentclass[letterpaper, 10 pt, conference]{ieeeconf}  

\usepackage{graphicx}
\usepackage{authblk}
\usepackage{cite}
\usepackage{amsmath}
\usepackage{siunitx}
\usepackage[font=normal,labelfont=bf]{caption}

\usepackage{array}
\newcolumntype{P}[1]{>{\centering\arraybackslash}p{#1}}
\newcolumntype{M}[1]{>{\centering\arraybackslash}m{#1}}

\IEEEoverridecommandlockouts                              
\title{\LARGE \bf The DAQ for the Single Phase DUNE Prototype at CERN }

\author{Roland Sipos$^{1}$ for the DUNE Collaboration%
\thanks{$^{1}$Roland Sipos is with CERN, CH-1211 Geneva 23, Switzerland}%
}

\begin{document}

\maketitle
\thispagestyle{empty}
\pagestyle{empty}

\begin{abstract}
DUNE will be the world's largest neutrino experiment due to take data
in 2025. Here is described the data acquisition (DAQ) system for
one of its prototypes - ProtoDUNE-SP due to take data in Q4 of 2018.
ProtoDUNE-SP also breaks records as the largest beam test experiment yet
constructed, and is a fundamental element of CERN's Neutrino Platform.
This renders ProtoDUNE-SP an experiment in its own right and the design
and construction have been chosen to meet this scale. Due to the
aggressive timescale, off-the-shelf electronics have been chosen to meet
the demands of the experiment where possible. The ProtoDUNE-SP cryostat
comprises two primary subdetectors - a single phase liquid argon time-projection chamber (LAr TPC) and
a companion Photon Detector. The TPC has two candidate readout solutions
under test in ProtoDUNE-SP - RCE (ATCA-based) and FELIX (PCIe-based).
Fermilab's artdaq is used as the dataflow software for the experiment.
Custom timing and trigger electronics and software are also described.
Compression and triggering will take the ~480\,Gb/s of data from the
front-end and reduce it sufficiently to 20\,Gb/s bandwidth to permanent
data storage in CERN's EOS infrastructure.
\end{abstract}

\section{INTRODUCTION}
Figure  1.  shows  the  ProtoDUNE-SP~\cite{Abi:2017aow}  TPC  volume  and  its  primary  components.  The  active  volume is  6\,m  high,  7\,m  wide  and  7.2\,m  deep  (along  the  drift  direction).  The  active  volume  consists of  a  central  cathode  plane  assembly  and  six  Anode  Plane  Assemblies  (APA).  Each  anode  plane  is constructed  of  three  adjacent APAs that  are  each  6\,m  high  by  2.3\,m  wide in  the  installed  position.  Each  APA  consists  of  a  frame  that  holds  three  parallel  planes  of  sense  and shielding  wires;  the  wires  of  each  plane  are  oriented  at  different  angles  with  respect  to  those  on  the other  planes  to  enable  3D  reconstruction.  The  wire  pitch  for  all  wire  planes  is  4.5\,mm,  and  each APA  holds  a  total  of  2,560  wires.  The  front-end  cold  electronics,  mounted  onto  the  APA  frame, and  immersed  in  LAr,  amplify  and  continuously  digitize  the  induced  signals  on  the  sense  wires  at 2\,MHz,  and  transmit  these  waveforms  to  the  Data  Acquisition  system  (DAQ).  From  the  DAQ the  data  are  transmitted  through  the  buffer  to  disk,  then  to  the  central  CERN  Tier-0  Computing Center,  and  finally  to  other  partner  sites  for  processing  and  analysis.
%
%
%

\begin{figure}[thpb]
\centering
\includegraphics[width=.5\textwidth]{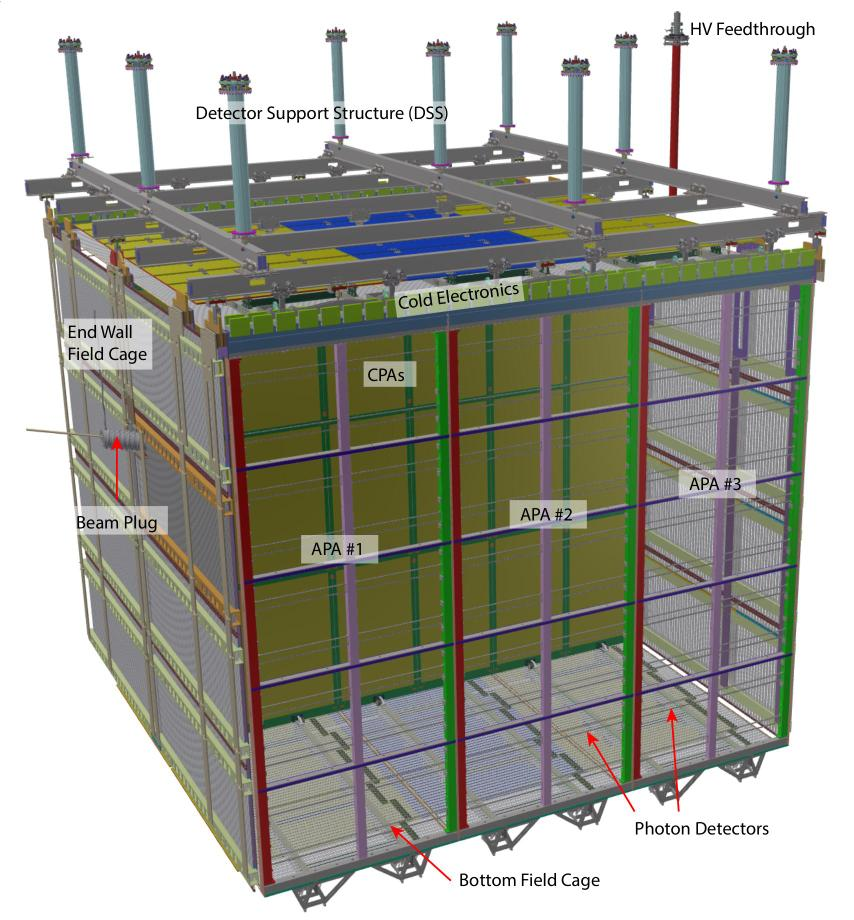} 
\caption{The particle detecting elements and major components of the ProtoDUNE-SP TPC.}
%
%

\label{PDSP_TPC}
\end{figure}

The readout of the TPC wires, prior to being received by the DAQ system, consists of cold electronics (CE) mounted on the APAs inside the cryostat and the warm electronics outside the cryostat on the flange. CE data are received on the Warm Interface Boards (WIBs) which are situated on the top of the flanges. Each WIB multiplexes the data to four 4.8\,Gb/s (or two 9.6\,Gb/s) lines that are sent over optical fibers to the DAQ. Two systems are used to receive data from the WIBs. The primary system is based on Reconfigurable Computing Elements (RCE)~\cite{Herbst:2016prn} which is used to read out 5 of 6 APAs, while the secondary system described here is based on the Front-End Link EXchange (FELIX)~\cite{Anderson:2016lfn} technology and is used to receive the data from the sixth APA.
%
%

\section{DETECTOR ENVIRONMENT}
ProtoDUNE-SP has 6 Anode Plane Assemblies (APAs), representing 4\% of a DUNE Far Detector Single Phase module. This yields 15,360 channels, that are continuously read out at 2\,MHz sampling rate. This results in ~450\,Gb/s data rate from the TPC. The  TPC  is  supported  by  a  Photon  Detector  System (PDS).  Each  PDS  module  consists  of  a  bar-shaped light  guide  and  a  wavelength-shifting  layer;  ten  of  which  are  mounted  inside  each  APA  frame. Each of them sampled at 150\,MHz.

The detector is located on the CERN SPS beam line, which allows for an exposure to charged particle beams in the momentum range 0.5-7.0 GeV/c.  Its surface location additionally exposes it to a high cosmic ray flux. The SPS super cycle structure is 2 times of 4.8\,s extractions in 32.4\,s.

Considering these environmental aspects, the DAQ system has strict requirements. It needs to be a single, scalable system across all sub-detectors, supporting different interfaces, without differentiation on detector type from operations point of view. Partitioning is a key concept to provide complete flexibility, in order to provide the possibility to read out different detector components at any combination in parallel. The DAQ will operate without dead-time under normal conditions, so continuous storage of overlapping ranges of data in space and time needs to be provided.

\section{THE DAQ SYSTEM}
Due to the extremely tight construction and commissioning schedule (data taking in Q4 2018), the DAQ is designed to use commercial off-the-shelf (COTS) components, hence most of the implemented solutions are based on existing frameworks and generic technologies.
%
%
Based on the previously described detector environment, the DAQ system also needs to be externally triggered, in order to limit the amount of data gathered. To provide this requirement, a custom board for implementing the trigger logic was developed, that accepts inputs from beam instrumentation, muon veto and photodetectors, and assembles a single global trigger to the system. The readout system also heavily relies on large buffers to allow the exploitation of the spill structure.

The general overview of the system is seen as follows, on figure 2.
\begin{figure}[thpb]
\centering
\includegraphics[width=.52\textwidth]{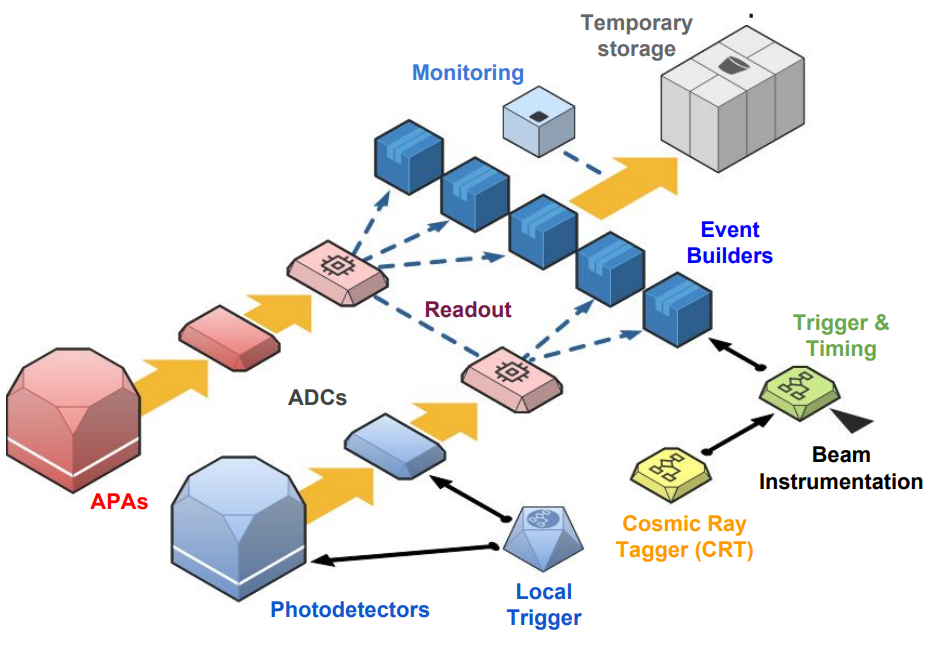} 
\caption{General overview on the DAQ system.}
\label{daq-overview}
\end{figure}
%
%
\subsection{DAQ in numbers}
In case of continuous readout, the averaged bandwidth over the SPS cycle is $\sim 135$\,Gb/s. With the expected DAQ compression ratio of 4, this results in an average storage throughput of $\approx34$\,Gb/s. 
The average data rate over the spill is expected to be about $3$\,Gb/s. An event is a $5$\,ms time window of all data contained in the readout, corresponding to the timestamp of a trigger, and results in $\sim 60$\,MB, if the x4 compression is achieved.

The DAQ  ensures proper decoupling between the “online” and the “offline” worlds, so sufficient temporary storage space for raw data files on the DAQ side are installed. The system can store up to 3 days worth of data taking, that results in the minimum of $135$\,TB ($25$\,Hz in spill + x4 compression). In order to have safe boundaries, $700$\,TB of usable disk space is installed.
The data is streamed to the CERN's OpenSource Storage (EOS)~\cite{eos}, with a maximum allowed throughput of $20$\,Gb/s.

The following table shows a summary of important numbers of system characteristics:
%
%
\begin{table}[h!]
\centering
\begin{tabular}{ |M{2.5cm}||M{2.5cm}|M{2.5cm}|  }
 \hline
   & Baseline & Maximum \\
 \hline
 \hline
 Trigger rate during spill & 25 Hz & 100 Hz \\
 \hline
 Average storage bandwidth & 17 Gb/s & 68 Gb/s  \\
 \hline
 Avg. bandwidth with x4 compression & 4.2 Gb/s & 17 Gb/s  \\
 \hline
 Peak & 56 Gb/s & 225 Gb/s  \\
 \hline
 Peak with x4 compression & 14 Gb/s & 56 Gb/s \\
 \hline
\end{tabular}
\end{table}

\subsection{Front-end electronics}
There are 2 different modules that represent the input for the DAQ system, one to read the TPC's APAs, and another to gather data from the photodetectors. The Warm Interface Boards (WIB) interface from cold electronics to the DAQ, with real-time diagnostic possibilities. The cold-electronics are the ProtoDUNE Front-end Motherboards (FEMBs), and a WIB multiplexes data from 4 of them. Each data stream sends $56$\,bytes per $500$\,ns, that results in $\approx 3.6$\,Gb/s payload per FEMB. The WIB is connected via optical links to the data acquisition and slow control systems. Each APA is isolated inside the cryostat and only connected to the detector ground through the cold-electronics at its own cold-electronics flange, where 5 WIBs are reading out the APA. This is seen on figure 3.
%
%

%
%

\begin{figure}[thpb]
\centering
\includegraphics[width=.4\textwidth]{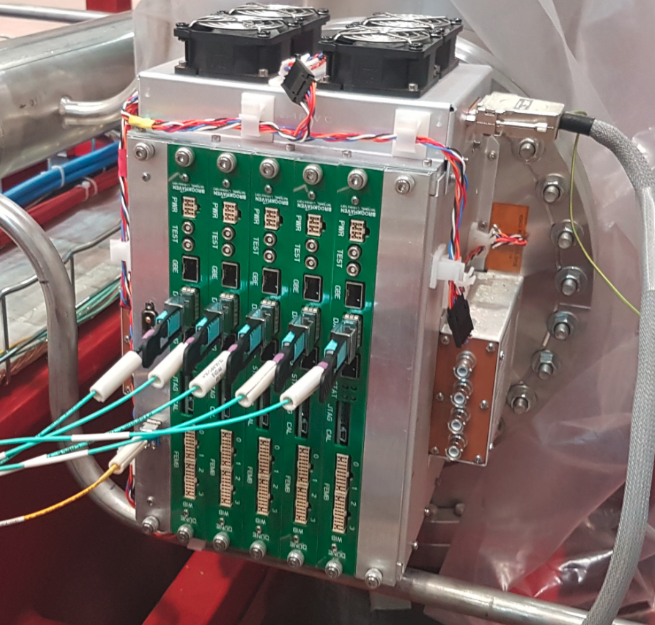} 
\caption{5 WIBs are connected to 1 APA per flange; all connections are optical for electrical isolation.}
%
%
\label{wibs}
\end{figure}

The readout system of the photodetectors is the Silicon Photomultiplier Signal Processor (SSP) prototype module, that is a high-speed waveform digitizer, with 12 channels per module. It has current sensitive, differential input amplifiers that provide good noise performance over long cables. The timing is obtained using signal processing techniques on the leading edge of SiPM signal (CFD), using the on-board Artix FPGA. It has deep data buffering (\SI{13}{\micro\sec}), and operates with no dead-time up to $30$\,KHz per channel.

\subsection{Timing system}
The timing system is required to: provide a stable and phase-aligned master clock to all DAQ components; receive external signals (including triggers) into the ProtoDUNE clock domain and time-stamp them; distribute synchronization, trigger and calibration commands to the DAQ system; and conduct continuous checks of its own function. In addition, the timing system acts as a data source, providing a record of timing signals received, distributed, or throttled. 

An FPGA-based master unit receives a high-quality clock (provided in ProtoDUNE by a GPS-disciplined oscillator) and external signals from the trigger system and SPS accelerator, as seen on figure 4.
\begin{figure}[thpb]
\centering
\includegraphics[width=.6\textwidth]{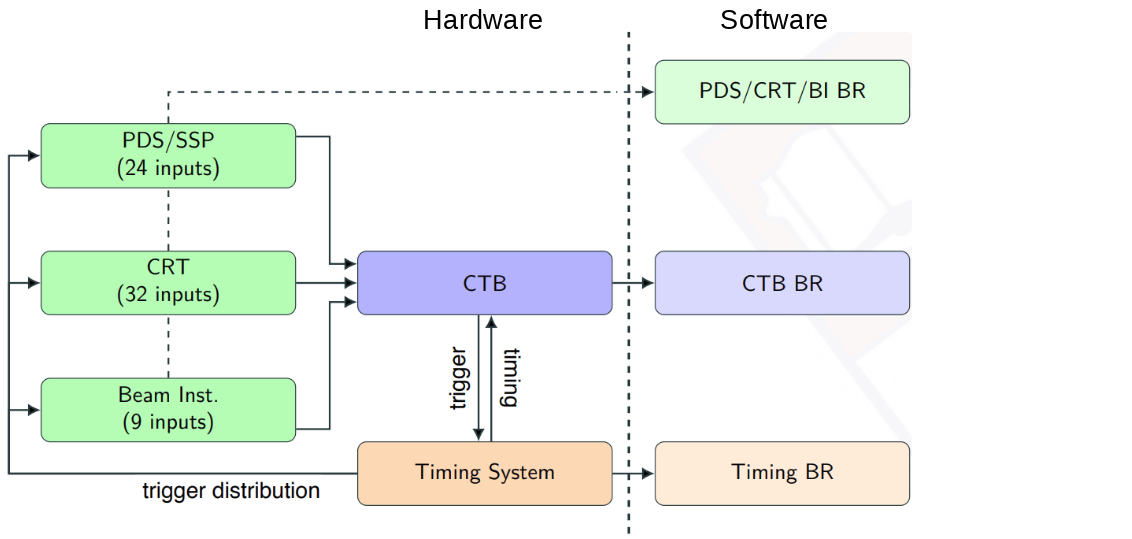} 
\caption{Timing system logic. }
%
%

\label{timing-ow}
\end{figure}

The master unit multiplexes synchronization and trigger commands, along with arbitrary command sequences generated by software, into a single encoded data stream, which is broadcast to all timing endpoints, and decoded into separate clock and data signals. The same high-quality clock is used for the beam instrumentation timing such that both
systems remain synchronized. A uniform phase-aligned cycle counter, updating at the ProtoDUNE system frequency of $50$\,MHz, is maintained at all endpoints, allowing commands to take effect simultaneously at all endpoints regardless of cable lengths or other phase delays. The system uses duplex links, allowing all endpoints to be regularly interrogated during system operation to verify correct operation and reception of timing commands.

\subsection{Trigger system}
The Central Trigger Board (CTB) is designed to serve as the master trigger, assembling information from the beam instrumentation, photodetectors, and the cosmic ray tagger, and to broadcast it to the timing network. The CTB is designed around the MicroZed System-on-Chip (SoC) board, equipped with a Xilinx Zynq7020. The motherboard implements the hardware interface with different systems, the FPGA implements trigger logic and interfaces with timing, finally the CPU and software elements manage the FPGA configuration and communication with DAQ software.

This solution supports up to 100 separate inputs which is deemed sufficient for ProtoDUNE with spares for each input type (fibre, BNC, etc.). The firmware is designed to be suitably flexible such that the exact trigger selection can be decided at configuration time, and does not require a new firmware. It will be able to decide on a trigger well under \SI{1}{\micro\sec}. The CTB is housed in the ProtoDUNE DAQ racks (seen on figure 5.) and electrically isolated from the cryostat where necessary (i.e., use of fibre). 
%
%

\begin{figure}[thpb]
\centering
\includegraphics[width=.4\textwidth]{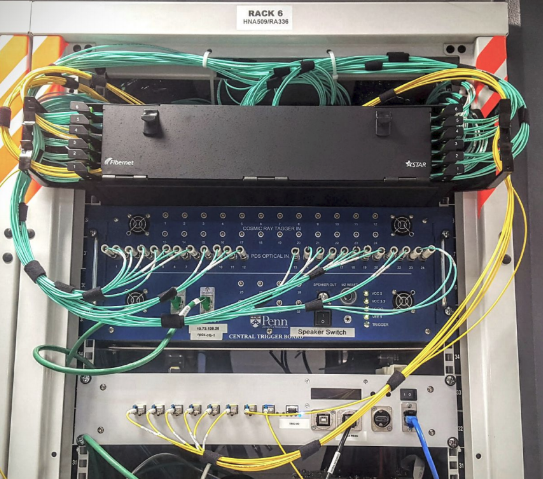} 
\caption{The Central Trigger Board in the DAQ rack.}
\label{ctb}
\end{figure}

The general outline and logic of the trigger system is seen on figure 6.
\begin{figure}[thpb]
\centering
\includegraphics[width=.5\textwidth]{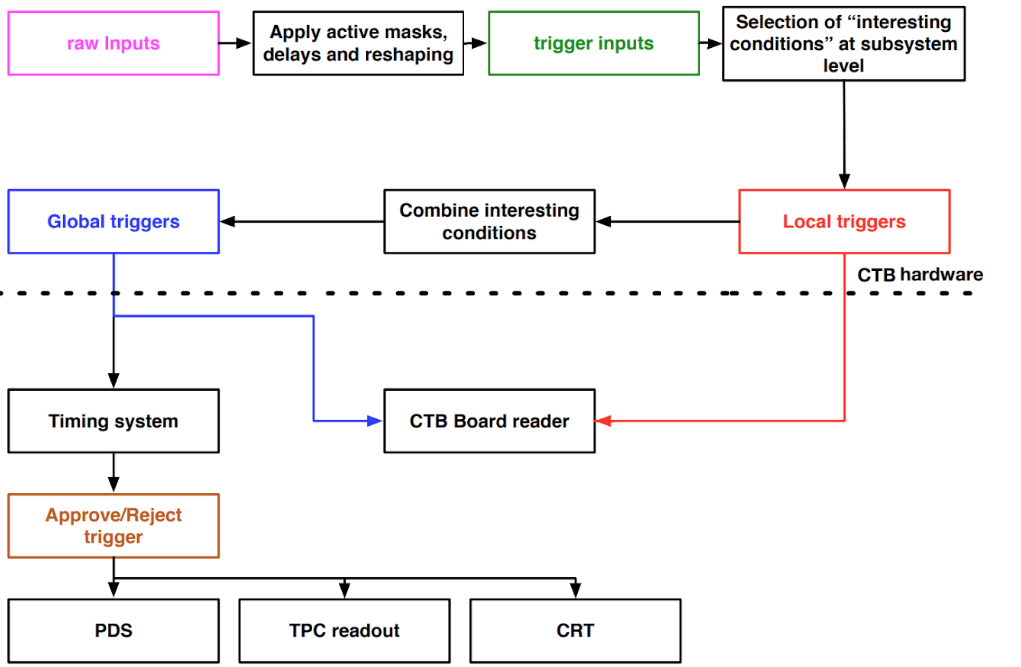} 
\caption{The Trigger system logic.}
%
%
\label{trigger}
\end{figure}
%
%
\section{READOUT SYSTEMS}
There are several flavors of off-detector readout systems in ProtoDUNE:
\begin{itemize}
\item The readout systems for Photon detectors as well as for the Trigger and Timing
system are fully software based.
%
%
%
\item For the TPC readout two solutions are being implemented, to be evaluated and
compared; the ATCA based SLAC solution (RCE) and the PCIe based FELIX
solution. The aim is to identify strong and weak points of both solutions.
\end{itemize}
For the TPC readout systems, there are two firmware variants that can be uploaded onto the WIBs FPGAs. For all readout systems the software application that interfaces to the event building farm is the so-called Board Reader. As the name suggests, this application type is tailored to read data from specific detector electronics, prepare event fragments and send them to the event building system.

\subsection{RCE readout}
The Reconfigurable Cluster Elements (RCE) based readout is a full meshed distributed architecture, based on networked SoC elements on ATCA platform, developed by the SLAC National Accelerator Laboratory. Being the baseline solution, it reads 5 APAs, a total of $12,800$ channels. The ATCA crate mounted in the DAQ racks is seen on figure 7.
\begin{figure}[thpb]
\centering
\includegraphics[width=.45\textwidth]{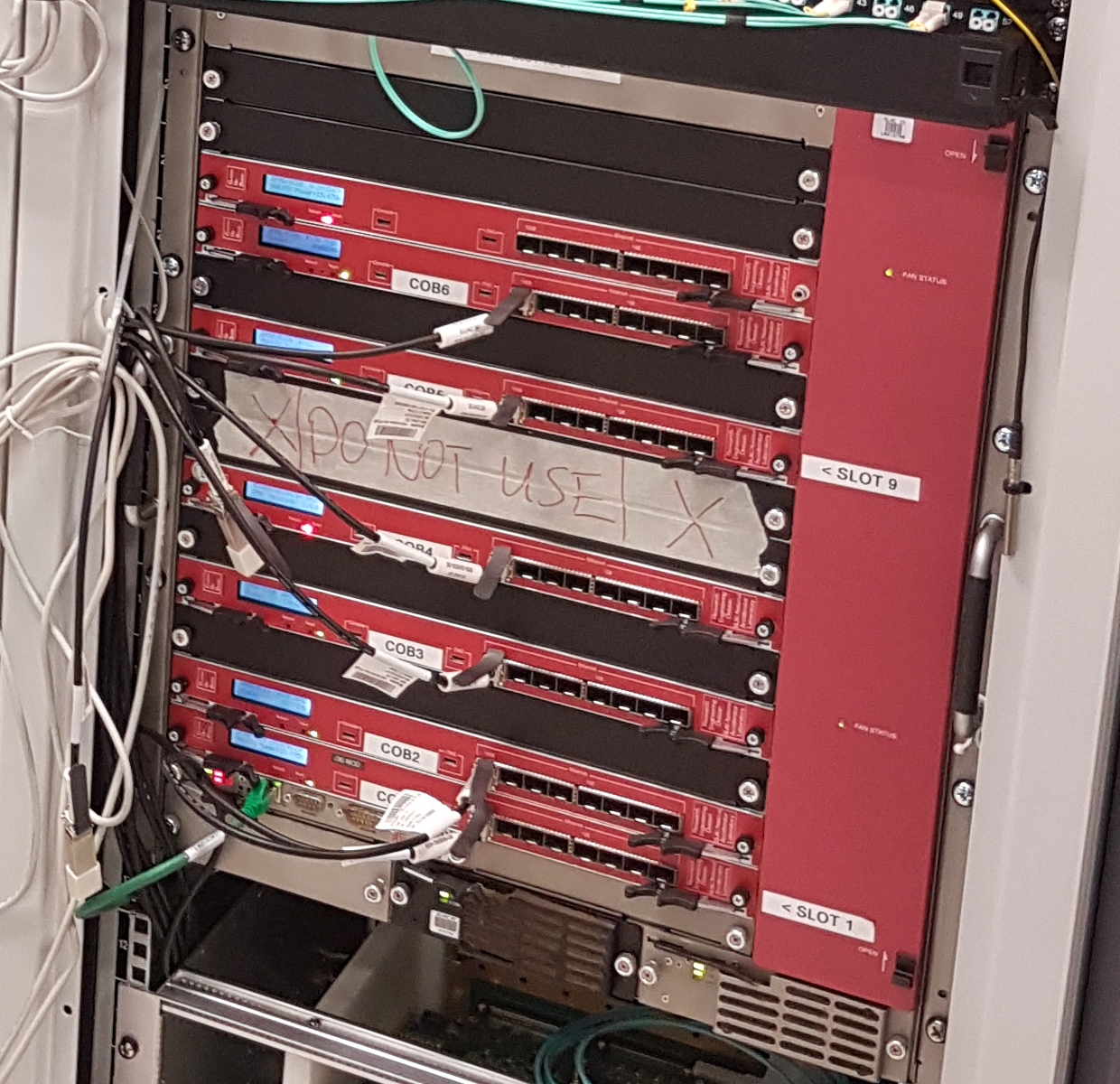} 
\caption{The RCE readout ATCA crate.}
\label{trigger}
\end{figure}

Each Cluster On Board (COB) hosts 9 processing elements, each with dual core ARM A9 processors and 1 GB DDR3 memory. The COB also carries a data processing daughter board with dual Zync 045 FPGAs. The RTM (Rear Transition Module) at the back of the COB supports the experiment specific interfaces, seen on figure 8.
\begin{figure}[thpb]
\centering
\includegraphics[width=.5\textwidth]{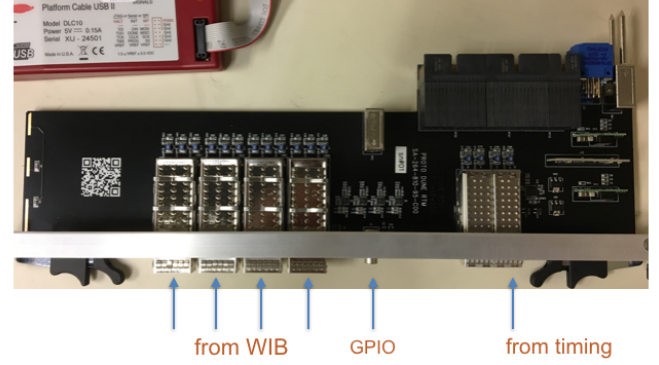} 
\caption{The ProtoDUNE specific RTM for the RCE readout.}
\label{trigger}
\end{figure}

This particular approach is focusing on early data processing, with tightly coupled custom firmware and software implementations. The ProtoDUNE version of the RCE will accept digitized data from the WIBs over optical fibre. The data are sent to a set of processing units where it is checked for errors, time-stamped, aggregated, formatted, compressed, buffered, selected, and then sent on the backend DAQ over TPC/IP upon receiving an external trigger. An output data rate $\sim 1$\,Gb/s per RCE can be sustained. A compression factor of 4 or above is expected (but contingent on noise). 

\subsection{FELIX readout}
A single APA (2560 channels) will be read out using the FELIX system. The Front-End Link EXchange (FELIX) is a project initially developed within the ATLAS Collaboration at CERN. Its purpose is to facilitate the development of high-bandwidth readout, needed for the High-Luminosity LHC, presently planned to start in 2026. The motivation for FELIX is the desire to move away from custom hardware at as early a stage as possible, and instead employ commercial off-the-shelf PC-based hardware and networking. The FELIX design is based on a shared firmware/software solution. A PCIe card is used to stream input data arriving from the detector front-ends to a circular memory buffer in a host PC using a continuous DMA transfer (with fixed 1\,kB block size); a publisher software running on the host PC routes the data to multiple output destinations using network interface cards. The BNL-711 card with Xilinx Kintex Ultrascale and 48 optical links (MiniPODs) is seen on figure 9.
\begin{figure}[thpb]
\centering
\includegraphics[width=.5\textwidth]{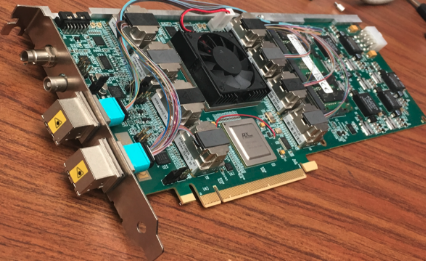} 
\caption{The FELIX BNL-711 card.}
\label{felix}
\end{figure}

For ProtoDUNE, CPU-based compression routines will be used in the FELIX design (in contrast to FPGA-based compression in the RCEs). This task can be also offloaded from the CPU, with the possible hardware acceleration, provided by Intel\textsuperscript{\textcopyright} QuickAssist (QAT)~\cite{qat} technology.

\section{DAQ SOFTWARE LAYER}
The entire software layer mostly depends on existing solutions and frameworks in order to meet the aggressive schedule.

\subsection{Data-flow and the artdaq framework}
The data-flow software for the ProtoDUNE SP DAQ is primarily responsible for acquiring the data from the readout electronics, packaging it appropriately, and storing it in files that are local to the DAQ cluster. It has responsibility for other functions that include the delivery of configuration parameters to the front-end electronics and real-time monitoring of the quality of the data and the performance of the DAQ system. It is not responsible for the transfer of the raw data files to permanent storage. The design of the ProtoDUNE DAQ data-flow software is based on artdaq~\cite{Biery:2013cda}, which is a data acquisition toolkit developed at Fermilab.

The different sub-processes are described as follows:
\begin{itemize}
\item {\bf BoardReader} - These processes are responsible for communicating with the detector electronics.
\item {\bf EventBuilder} - These processes are responsible for assembling complete events and optionally processing them in art.
\item {\bf DataLogger} - These processes are typically responsible for logging the data and serving events to real-time data quality monitoring (DQM) processes.
\item {\bf DataFlow Orchestrator} - It is responsible for queuing triggered events and load-balancing them among the EventBuilders.
\end{itemize}
The underlying hardware will be fully based on COTS servers running Linux CentOS7 and on a $10$\,Gb/s network (single device). Overall, 10-15 servers are sufficient to implement the complete data-flow for ProtoDUNE (board readers and data flow orchestrator, event builder, storage).

Figure 10. shows a possible data-flow based on artdaq.
\begin{figure}[thpb]
\centering
\includegraphics[width=.51\textwidth]{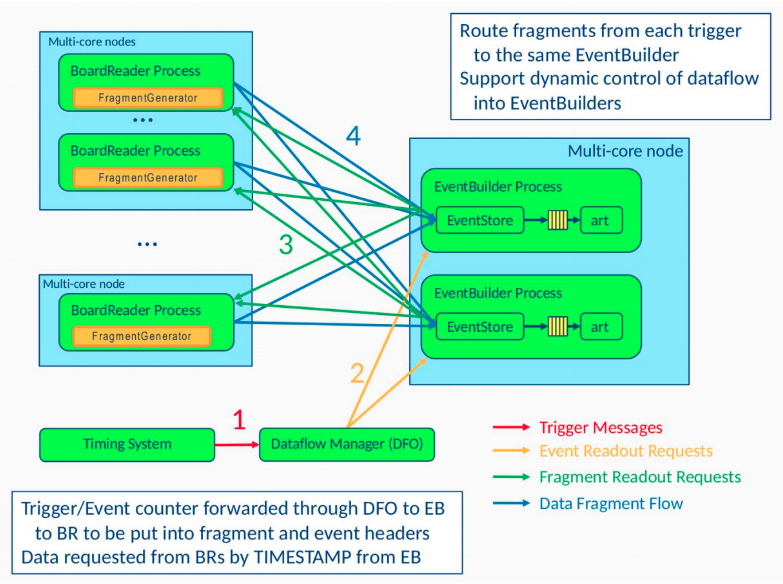} 
\caption{Data-flow in the artdaq framework.  }
%
%
\label{artdaq}
\end{figure}

Specific applications have additional functionalities:  
\subsubsection{SSP BoardReader}
The SSP hardware is responsible for recording, and making available to downstream components, waveforms from the photodetectors. The SSP is capable of triggering internally on photon signals, on an external trigger source, and at fixed time intervals, and can also generate trigger signals for other detector components. Using an on-board FPGA, it can calculate metadata related to the observed pulses (heights,widths etc.), which may also be sent in lieu of the actual waveform data in order to reduce data volumes. In turn, the BoardReader program is responsible for receiving data from an SSP unit, packaging it into fragments in time corresponding to those from the other detector systems, and sending it on to the EventBuilder to produce full events. The SSP BoardReader also does the trigger selection, it takes the trigger request from the EventBuilder and assembles a $5$\,ms window of any internal trigger activity around the external trigger. It also deals with configuration of the SSP hardware and deals with state transitions to/from the control system.
%
%

\subsubsection{Timing and trigger BoardReaders}
Besides their specific function on the system, the Timing and Trigger systems will also provide event data. The Timing system will provide the timestamp information of accepted triggers, dead-time information, etc. The Trigger will provide information about the specific inputs that contributed to trigger decisions. The information from those systems will be dealt with by dedicated BoardReaders, to be folded into the ProtoDUNE SP events, and in addition may be continuously gathered, irrespective of the trigger decisions.

\subsubsection{RCE BoardReader}
As the multiplexed data from the WIB come into the RCE FPGA fabric, it is de-multiplexed and buffered into per-channel, fixed-time-length chunks (for instance 512- or 1024-ticks). These chunks are compressed and written to the DRAM where the RCE processor waits for a trigger (also handled by the FPGA) to arrive. Upon a trigger, the processor sends data for a fixed window in time, including pre- and post-trigger time chunks for all channels, to the RCE BoardReader applications.

\subsubsection{FELIX BoardReader}
The BoardReader implementation for the FELIX based readout integrates a network messaging layer that subscribes to the FELIX publisher application. The implementation also focuses on flexibility, as the topology of the queues and links is scalable by the BoardReader configuration. In order to achieve the required performance, particular care has been put into the implementation, avoiding dynamic memory allocation. Internal elements and functionalities strictly avoid memory copies. Every link has dedicated subscriber threads that populate single producer single consumer queues, using lock-free implementation. Multi-threading features ensure the proper synchronization of the trigger matching threads, and also to comply with the internal state machine of artdaq. The compression facility relies on software techniques, with possible hardware acceleration.

\subsection{Run Control}
The DAQ uses a collection of tools currently in use for accomplishing its run control, operational monitoring, and configuration management. The run control software for the experiment must:
\begin{enumerate}
\item provide process control functionalities; launch and terminate DAQ applications,
\item provide a finite-state-machine (FSM), to ensure the proper sequence of process control steps,
\item be the interface to operators,
\item provide support to include, exclude and partition a given part of the system
\end{enumerate}

It is based on the Joint COntrols Project (JCOP)~\cite{jcop} extension for the WinCC-OA\textsuperscript{\textcopyright} supervisory control framework from Siemens\textsuperscript{\textcopyright}. It is in common use at LHC experiments, and it is officially supported by CERN. The JCOP system, interfaced to artdaq system, satisfies the above requirements and carries with it the added benefit of expertise and support available at CERN. Monitoring metrics for process statuses, errors, alerts, and trends is developed with such system for monitoring both the data-flow and detector status. JCOP is also used for the Slow Control system, providing a uniform interface to detector operation. The run configuration utilities for the DAQ provide
\begin{enumerate}
\item configurations for DAQ system architecture (describing the data flow), DAQ hardware, and DAQ software,
\item run numbers and assigning a user-defined configuration to that run,
\item partitioning support,
\item a UI for configuration generation/viewing,
\item a means to lookup previous run configurations.
\end{enumerate}
%
%

\subsection{Online monitoring}
A fraction of the data taken is also streamed to the online monitoring computers. The online monitoring functions to provide a quality check on the data. It is designed for rapid feedback and processing of an event should take no more than O(10) seconds. Higher level reconstruction (which can take minutes) is deemed for offline data quality. A non-exhaustive list of the types of quantities to be measured are TPC noise, photon dark/cosmic count, rate and types of triggers, error counts, hit maps, etc.
The Online Monitoring has a defined data-flow as follows:
\begin{enumerate}
\item DataLogger sends events to one or more dispatcher processes
\item Dispatcher processes route events to the online monitoring processes 
\item RAW decoder processes unpack the raw data to perform low-level analysis
\item ArtAnalyzer processes perform high-level analysis of the unpacked events
\item Histograms are saved and propagated to the web display
\end{enumerate}

The described flow is seen on figure 11.
\begin{figure}[thpb]
\centering
\includegraphics[width=.5\textwidth]{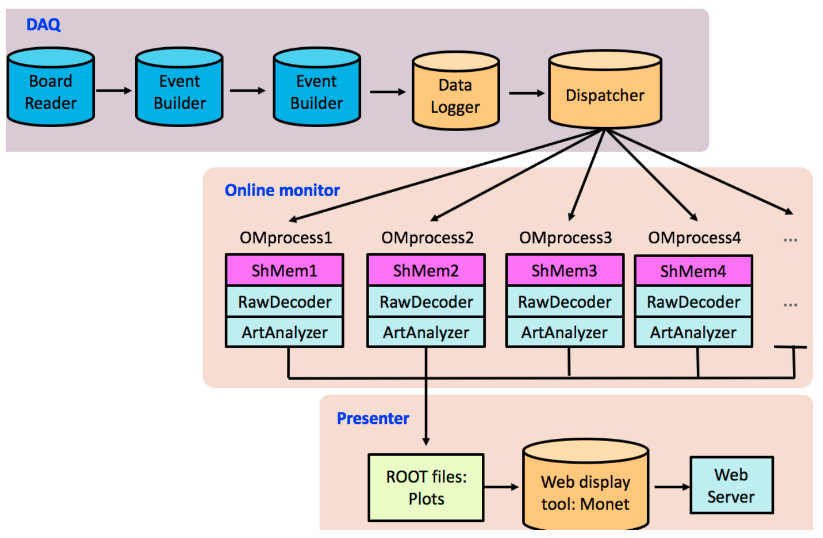} 
\caption{Data-flow in the Online Monitoring system.}
\label{om}
\end{figure}

\subsection{Backpressure}
The DAQ system is dimensioned to be able to sustain the requested rate of $25$\,Hz during an
SPS spill and is capable of making use of the inter-spill time to absorb any backlog that may
have built up during the spill.
Three different sources of backpressure have been identified for protoDUNE. In this section
they are described, together with an indication of the time scales at which every type of
backpressure builds up.
Backpressure is not expected at the level of the on/near detector readout electronics, since
the system has been dimensioned such that for the TPC (and for the SSP), data will be
read-out continuously (based on self triggering).

\subsubsection{Trigger rate spikes}
Triggering with beam is subject to variations in rate that may cause for an over-subscription
of DAQ resources. In order to prevent this from happening, artificial dead-time will be
managed inside the trigger board logics. As an example, simple dead-time logics will be
introduced, excluding the firing of the trigger at an interval closer than a defined fraction of
the readout window (~5 ms); additional complex dead-time logic in the trigger board may
forbid accepting more than a fixed number of triggers during a time interval (e.g. max 200
triggers during spill, max 50 triggers during inter-spill; at 25Hz over a 5s spill, 125 events are
expected on average per spill.

\subsubsection{Transient noise bursts and Event building/aggregation limitations}
Transient noise bursts may blow-up the size of data corresponding to a trigger. The limited
internal bandwidth in the RCE based readout system may cause a backlog of data to build
up. In order to prevent data loss, a BUSY software message may be sent towards the
trigger, when buffer occupancy reaches a certain threshold. When reaching a comfortable buffer occupancy, the BUSY shall be cleared. Temporary failure of one or more dataflow
components (a network link, an event builder node/process, an aggregator, a pool of disks,
etc...) may cause the data acquisition performance to be degraded. This will be noticed by
the data flow aggregator (lacking EB nodes to assign event to). Similarly to noise bursts, a
BUSY message shall be issued once queue lengths exceed a defined value.

\subsubsection{Storage limitations}
The DAQ writes raw data to a temporary storage area, waiting for the data to be transferred
to EOS. A long-term outage of the connection to EOS or its performance degradation may
cause the storage area to progressively fill up. This type of issue builds up in hours/days. No
transient BUSY mechanism can be applied in this case.
From the analysis of back-pressure sources it is apparent that in ProtoDUNE there is no
requirement for a fast feed-back to the trigger system to stop the generation of triggers.
Therefore a software based messaging system is well suited for throttling the trigger.
Nevertheless, provisions are made at the timing master level to be able to introduce a
hardware signal indicating busy conditions, in case it turned out to be needed.

\section{Summary}
We developed the ProtoDUNE Single Phase data acquisition system, that meets the requirements of challenging conditions and described constraints. In order to meet the extremely tight construction and commissioning schedule, most of the implementations are based on COTS hardware solutions and mature software tools and frameworks. As the DAQ needs to be externally triggered, a custom timing and trigger system was developed, that aggregates inputs from beam instrumentation, muon veto and photodetectors, and assembles a single global trigger for the system. The system provides two TPC readout solutions, in order to identify strong and weak points of each one. The DAQ is designed to handle the effects of backpressure, trigger rate spikes and transient noise bursts, the data flow orchestrator provide features to handle event building and aggregation limitations at higher trigger rates. ProtoDUNE-SP due to take data in Q4 of 2018, and being the largest beam test experiment yet constructed. The DAQ system faces a wide range of challenges, and it is ready for the beam run to prove its design principles.

\addtolength{\textheight}{-12cm}   


\section*{ACKNOWLEDGMENT}
We would like to express our gratitude to groups external to DUNE, who contributed to the development of the data acquisition system; CERN IT for the support on the DAQ servers and network, the FNAL artdaq team for providing the software framework, and the JCOP team for the run control and operational monitoring support.
We gratefully acknowledge the ATLAS collaboration for the assistance and development on the FELIX readout.
Last but not least to the Neutrino Platform for the infrastructural support at EHN1.
%
%

\bibliography{mybib}{}
\bibliographystyle{plain}

\end{document}